# Surface phonons and possible structural phase transition in a topological semimetal PbTaSe$_2$

Vivek Kumar[1,*] and Pradeep Kumar[1,†]

[1] School of Physical Sciences, Indian Institute of Technology Mandi, Mandi-175005, India

**Abstract**

Topological insulators are a novel class of quantum materials characterized by protected gapless surface or edge states but insulating bulk states which is due to presence of spin-orbit interactions and time-reversal symmetry. Such an intriguing surface and bulk topology manifests itself in coupling with lattice dynamics due to electron-phonon scattering. Here we report an in-depth investigation of a topological nodal line semimetal PbTaSe$_2$ via temperature, polarization dependent Raman spectroscopy and temperature dependent single crystal X-ray diffraction (SC-XRD) measurements. Our analysis shows signature of electron-phonon coupling as reflected in the Fano asymmetry in line shape of M1-M4 modes and anomalous temperature variation of linewidth of P3-P4 modes. Further polarization dependent phonon symmetry changes at different temperature (6K and 300K), discontinuities in bulk phonon dynamics for P2-P5 modes and disappearance of phonon modes i.e., M1-M5, on decreasing temperatures indicates towards a thermally induced structural phase transition which is also supported by the SC-XRD results. Hence based on our findings we propose that M1-M4 modes are surface phonon modes, the material undergoes a thermally induced structural phase transition from α to β phase at T$_{α→β}$ ~ 150 K or is in close proximity to the β phase and another transition below T$_{CDW+β}$ ~ 100K which is possibly due to the interplay of remanent completely commensurate charge density wave (CCDW) of 1H-TaSe$_2$ and β phase.

*vivekvke@gmail.com
†pkumar@iitmandi.ac.in

# 1. Introduction

The discovery of topological insulators (TIs) has prompted a deep interest of condensed-matter physicists towards finding new methods to generate exotic quantum phases in novel quantum materials. Other than their rich theoretical realm, topological semimetals got promising spintronic applications in magnetic nano-devices owing to the presence of spin-momentum-locked surface states and quantum computing [1]. Presence of Majorana fermions, symmetry protected robust metallic edge and surface modes with insulating bulk ground states has been important features of topological materials [2]. After the recent emergence of Weyl semimetals an imperative shift in focus towards Topological semimetals has been observed because of such promising properties and applications [2-9].

In the league of such materials, $PbTaSe_2$ has attracted interest of the condensed matter community. R. Eppinga et al., in 1980s reported it as one of the prepared intercalated compounds of the post-transition elements i.e., In, Sn, Pb and Bi with $TX_2$ (T= Ta and Nb; X = S and Se) [10]. It is a spin-orbit coupled non-centrosymmetric superconducting topological semimetal and has been reported to possess a nonzero $Z_2$ topology with fully spin-polarized Dirac surface state and topological nodal-line fermions [3,11-14]. D. Multer et al., reported evidences of robust nature of topological surface states at Fermi energy against dilute magnetic impurities [15]. M. N. Ali et al., found an existence of a gapped graphene-like, but heavier Dirac cone at K point in the electronic structure which is extracted from the hexagonal Pb layer and generates 3D massive Dirac fermions. Strong spin-orbit coupling which is inherent to heavy Pb atom, introduces a gap of 0.8 eV which can contribute to large Rashba splitting. Absence of inversion symmetry is a crucial component in lifting spin degeneracy of electronic bands and formation of topological nodal lines near Fermi level [16] . The topological character of the electronic structure has been identified in experimental and first principal Density Functional Theory (DFT) calculations. Transport and magnetization measurements indicated a metallic behavior until it reaches $T_c$ ~ 3.79 K below which it exhibits a type-II Bardeen-Cooper-Schrieffer (BCS) superconductor like character at ambient pressures [3,16-18]. Interestingly, the structure of stochiometric PbTaSe$_2$ has been found to be extremely sensitive to the pressure. It undergoes a series of transition from P$\bar{6}$m2 (α-phase) to P6$_3$mc (β-phase) at sub-GPa to P6/mmm (γ-phase) at ~7.5 GPa and finally to Pmmm (δ-phase) at ~ 44 GPa successively [19-22]. In ambient surrounding conditions XRD pattern revealed that PbTaSe$_2$ crystallizes into a hexagonal crystal structure with space group P$\bar{6}$m2 (α-phase, #187, $D_{3h}$) and extracted lattice constants are $a = b = 3.45 \overset{0}{\text{A}}$, $c = 9.35 \overset{0}{\text{A}}$ and $\alpha = \beta = 90^o$ and $\gamma = 120^o$. In this phase, an alternate hexagonal stacking of TaSe$_2$ and Pb layers is present, position of Ta atom is such that it breaks inversion symmetry as shown in Fig. 1(a,b). Pb, Ta and Se atom occupy 1a, 1d and 2g Wyckoff positions, respectively [3,21].

Charge density waves (CDWs) are periodic spatial modulations of the electronic charge density in certain materials. CDWs originates from the coupling of electronic structures with the lattice distortions which was initially postulated by Peierls and later by Fröhlich in order to characterize metal-insulator transition [23]. Transition metal dichalcogenides (TMDs) are known to sustain CDWs and their applicability in optoelectronics and energy storage devices. Bulk 2H-TaSe$_2$ has shown presence of CDW instability below ~122 K where it undergoes from metallic to incommensurate CDW (ICDW) followed by a completely-commensurate CDW (CCDW) below ~90K [24-28]. The conundrum of competition between superconductivity and CDWs is well known, for example in case of compressed topological Kagome metal X (Cs, Rb) V$_3$Sb$_5$, Cu$_x$TiSe$_2$ and TMDs like NbSe$_2$ and 1T-TaSe$_{2-x}$Te$_x$ [29-34].

Phonons plays a vital role when it comes to determining lattice and electronic transport behavior of materials. Raman spectroscopy is the phenomenon of light-matter interactions which can provide information of underlying electron-photon, electron-phonon, electron-electron interactions etc. [35,36]. The presence of strong electron-phonon coupling in topological semi-metals like TaAs, NbAs and WTe$_2$ has also been investigated using Raman spectroscopy [35,37-39] . It is sensitive to the crystal symmetry of the material as different vibrational modes, including surface phonons, will exhibit specific Raman activity based on the selection criteria associated with the crystal symmetry. Surface phonons are localized near the boundary or surface of a material. Their behavior is influenced by the disruption of the crystal lattice at the surface. Surface phonons can substantially affect the thermal conductivity, electronic transport, and other material properties, particularly in nanoscale systems where the surface-to-volume ratio is high. Raman spectroscopy has proved to be an excellent non-destructive probe to investigate the presence of CDWs [26,40], superconductivity [41,42] and surface phonons [43,44].

It has been theoretically predicted from the phonon spectrum that superconductivity arises as a consequence of introduction of Pb atom in TaSe$_2$ lattice [12]. Still, how underlying CDW phase, superconductivity and topological nature interplays with evolution of temperature in PbTaSe$_2$ has not been pondered upon much. Hence, motivated by these facts we decided to investigate this stoichiometric compound via Raman spectroscopic technique supported by DFT calculation and SC-XRD measurements.

## 2. Results and Observations

### 2.1 Group Theory

*2H*-TaSe$_2$ at high temperatures possess a hexagonal P6$_3$/mmc ($D_{6h}^4$) symmetry and the irreducible representation of Raman active modes is given as $\Gamma_{Raman}$ = A$_{1g}$ + 2E$_{2g}$ + E$_{1g}$ [28]. PbTaSe$_2$ in P$\bar{6}$m2 (α-phase, #187, D3h) has an effective of 4 atoms per unit cell as also clear from Figure 1(a). Group theoretical prediction of phonon modes at Γ-point is given as $\Gamma = A_1' + 3A_2'' + 3E' + E''$ ($\Gamma_{acoustic} = A_2'' + E'$ and $\Gamma_{optical} = A_1' + 2A_2'' + 2E' + E''$), out of which Raman active modes are $\Gamma_{Raman} = A_1' + 3E' + E''$ and $\Gamma_{Infrared} = 3A_2'' + 3E'$ are infrared active modes [45,46]. $A_1'$ is symmetric with respect to (w.r.t.) principle-axis and reflection in the horizontal plane of symmetry; $A_2''$ is symmetric w.r.t principle-axis and anti-symmetric reflection in the horizontal plane of symmetry. $E'$ and $E''$ both are doubly degenerate modes where former is symmetric and later is anti-symmetric to the reflection in horizontal plane of symmetry. Table-I summarizes Γ-point phonon mode decomposition for respective Wyckoff sites along with Raman tensors for α-phase. The character table for D$_{3h}$(P$\bar{6}$m2) is summarized in Table-SI in Supplementary Information (S.I.) [47].

## 2.2 Experimental and Computational details

We have performed temperature dependent SC-XRD and Raman Spectroscopy to study the material. Temperature dependent SC-XRD has been carried out at ambient pressure to investigate the structural stability using SuperNova X-ray diffractometer system with copper (Cu, λ ~ 1.54 $\mathring{A}$) as a source. Liquid nitrogen is utilized to vary temperature from 87K till room temperature. Our room temperature SC-XRD data matches very well with the earlier report by Raman Sankar et al. [18]. We found appearance of new peaks on decreasing the temperature which suggests a structural transition. Further details are mentioned in SI [47].

Here, we will focus on the results of Raman spectroscopy where we captured inelastically scattered light via micro- Raman spectrometer (LabRAM HR Evolution) in the backscattering configuration. A 633nm (1.96 eV) linearly polarized solid-state LASER is used to illuminate the sample which is focused via 50X long working distance (LWD) objective lens with a N.A. of 0.8.

Laser power was maintained low (< 0.5 mW) to prevent local thermal effects. A Peltier cooled charge-coupled device (CCD) detector collected the scattered light after getting dispersed by a 600 groves/mm grating. Sample is inserted inside a high vacuum chamber with pressure reduced to ~ 0.1 mPa. The sample temperature is modulated over a range of 6K-320 K with an accuracy of ± 0.1 K using a closed-cycle He-flow cryostat (Montana). We have also performed a polarization dependent study to probe the symmetry of the phonon modes. Here we have done the measurement in a configuration where the incident light polarization direction is rotated keeping the analyzer fixed at 14K and room temperature (~ 300K). A pictorial representation of plane projection of this configuration is shown in Fig. 1 (c).

Structural optimization and Zone-centered phonon frequencies were calculated utilizing plane-wave approach as implemented in QUANTUM ESPRESSO [48]. Linear response method within Density-Functional Perturbation Theory (DFPT) is used to get dynamical matrix. Projector-Augmented Wave (PAW) pseudopotentials and Perdew-Burke-Ernzerhof (PBE) is used as an exchange-correlation functional. The kinetic-energy and charge-density cutoff is taken 40 Ry and 320 Ry respectively. The Monkhorst-pack scheme with 30 x 30 x 30 k-point dense mesh is used for the numerical integration of the Brillouin Zone (BZ). Electron-phonon interaction is calculated by interpolation over the BZ as reported by M. Wierzbowska et al. [49]. Obtained phonon frequencies, optical activity and corresponding electron-phonon interaction energies are tabulated in Table SII (S.I.). Frequencies are found close to the experimentally observed modes and calculated mode symmetry is in agreement with the group theory prediction for α-phase as well.

## 2.3 Temperature evolution of the phonon modes

Study of temperature dependence of the Raman spectra can provide a vital information about anharmonicity, electron-phonon coupling of respective phonon mode, phase transitions and presence of other quasi-particle excitations. We observed five modes at lowest recorded temperature 6K labeled as P1-P5 for convenience. P2-P5 modes are consistent throughout the temperature range of 6K-320K. P2-P3 mode shows a $E'(x)$ or $A'_1$ kind of symmetry at 14K (refer section 2.5 for further discussion). In addition to these, some weak modes appear in certain temperature range only and are labeled as M1-M5, see in Fig. 2 which shows temperature evolution of the raw Raman spectra. M1-M4 modes exhibit asymmetry in the line shape which is indicated by dotted black line. We have fitted the Raman spectra at different temperatures with Lorentzian function as shown in Fig. S3 (S.I.) [47] and extracted corresponding variation of phonon features (frequency, linewidth and intensity) with temperature. Experimentally observed mode frequencies are mentioned in Table II.

A non-linear temperature dependence of the phonon features i.e., Frequency, full-width half maxima (FWHM) may be comprehended via a semi-quantitative model. According to this paradigm there is a combined effect of anharmonicity and thermal expansion. Anharmonic effect arises due to phonon-phonon interaction and is reflected in change in self-energy parameters whereas thermal expansion is a pure volume effect. It can be expressed in temperature variation of phonon frequency as follows:

$$\omega(T) = \omega_o + \delta\omega_{an} + \delta\omega_E \qquad \text{--(1)}$$

here $\omega_o$ is frequency at $T = 0K$, $\delta\omega_{an}$ and $\delta\omega_E$ is the contribution of anharmonic and lattice thermal expansion respectively [50,51]. Analysis on the effect of thermal expansion is mentioned in the section SI of S.I. [47].

Here we will discuss the effect of anharmonicity on frequency, FWHM and Intensity that comes into picture due to presence of anharmonic terms in the lattice potential. To comprehend the microscopic origin of anharmonicity in case of periodic solids one can think of atoms vibrating about their equilibrium position around which potential energy can be Taylor expanded as [52,53]:

$$U(q) = U(q_o) + q \left.\frac{\partial U}{\partial q}\right|_{q=q_o} + q^2 \left.\frac{\partial^2 U}{\partial q^2}\right|_{q=q_o} + q^3 \left.\frac{\partial^3 U}{\partial q^3}\right|_{q=q_o} + ... \qquad -- (2)$$

or equivalently: $\qquad U = const. + U_{harmonic} + U_{anharmonic} \qquad\qquad -- (3)$

$$U_{anharmonic} = gq^3 + mq^4 + ... \equiv \Pi\ (a^+a^+a + a^+aa) + \Lambda\ (a^+a^+a^+a + a^+a^+aa + ..) + ... \qquad -- (4)$$

where '$a^+$' and '$a$' are the creation and annihilation operators, $q$ is the normalized co-ordinate. Keeping energy and wavevector conservation in consideration, cubic term $g\ q^3$ *or* $\Pi\ (a^+a^+a + a^+aa)$ reflects the three-phonon process where an optical phonon mode decay into two equal energy acoustic phonon modes $(\omega_1 = \omega_2 = \omega/2; k_1 + k_2 = 0)$. While quartic term $m\ q^4$ *or* $\Lambda\ (a^+a^+a^+a + a^+a^+aa + ..)$ indicates four-phonon process where optical phonon decomposes into three acoustic phonons $(\omega_1 = \omega_2 = \omega_2 = \omega/3; k_1 + k_2 + k_3 = 0)$. Coefficients *g or* $\Pi$ and *m or* $\Lambda$ and $\Pi$ phenomenologically indicates the strength or contribution of three and four phonon process in anharmonicity. At low temperatures three phonon process is dominating whereas a minor contribution from four phonon. Temperature dependence of the phonon frequencies having the contribution of both three and four-phonon process can be described using following functional form [54]:

$$\delta\omega_{an}(T) = \omega(T) - \omega_o = A\left(1 + \frac{2}{e^x - 1}\right) + B\left(1 + \frac{3}{e^y - 1} + \frac{3}{(e^y - 1)^2}\right) ; \quad \text{-- (5)}$$

Whereas the FWHM, $\Gamma_{an}(T)$, can be expressed as following [55]:

$$\Gamma_{an}(T) = \Gamma_{an}(0) + C\left(1 + \frac{2}{e^x - 1}\right) + D\left(1 + \frac{3}{e^y - 1} + \frac{3}{(e^y - 1)^2}\right) \quad \text{-- (6)}$$

respectively, here $\omega_o$ and $\Gamma_{an}(0)$ is mode frequency and line-width at absolute zero temperature; $x = \frac{\hbar\omega_o}{2k_B T}$, $y = \frac{\hbar\omega_o}{3k_B T}$ and $k_B$ is the Boltzmann constant. Coefficient A/B and C/D represents the strength of phonon-phonon interaction involving three/four phonon process, respectively. Higher order terms can be neglected as they start contributing at much higher temperatures.

Temperature evolution of the frequency, FWHM and Intensity of some of the prominent phonon modes P2, P3, P4 and P5 are shown in Fig. 3 (solid orange line is a guide to eye, (dashed line shows position of discontinuity). We note that in case of frequency only P2 and P3 shows a significant anharmonic effect so we have fitted these with eq. 5 in temperature range 100K-320K and fitted parameters are mentioned in Table II. Fitted curve is shown in thick red line and solid light blue line is a linear extrapolation which reflects the deviation from the anharmonic prediction. On decreasing the temperature both modes i.e., P2 and P3, frequency blueshifts in temperature range from 320K to ~100K and shows significant softening below 100K till 6K. P4 mode frequency show normal behavior from 320K to ~220K i.e., mode frequency increases with decrease in temperature. From ~ 220K to ~120K it shows a sharp decrease and thereafter it remains nearly constant till 6K. Similar anomalous behavior is seen in P5 where it blueshifts almost linearly till ~200K and on further decreasing the temperature it monotonously redshifts till 6K. On decreasing the temperature, the linewidth of P2 mode increases linearly from 320K to ~200K and then decreases continuously till 6K. In case of linewidth of P3 and P4 mode an intriguing behavior is observed that, other than the discontinuity at ~100K and 150K respectively, it monotonously increases with decreasing temperatures. Such an anomalous nature indicates a possible presence of electron-phonon coupling which is discussed in the next section i.e., 2.4. For mode P5 linewidth shows anomalous behavior i.e., increases with decreasing temperature. Now we will discuss the temperature evolution of the intensity of these modes (P2-P5). We know that generally bosonic (phonons) population decreases with decrease in temperature. P2 shows an expected behavior as its intensity decreases with decrease in the temperature. An interesting behavior is observed in case of P3 where the intensity decreases from 320K to ~150K and increases on further decreasing the temperature till 6K. P4 shows anomalous behavior where it remains almost constant throughout the temperature range. For P5 intensity remains constant from 320K to ~150 K and then increases monotonously on further decreasing the temperature which is anomalous as well.

Interestingly we observed very strange temperature dependent nature of the modes labeled as M1 (~ 60 $cm^{-1}$), M2 (~ 77 $cm^{-1}$), M3 (~ 165 $cm^{-1}$), M4 (~ 182 $cm^{-1}$) and M5 (~ 235 $cm^{-1}$). M1, M2 and M3 disappears below ~100K whereas M4 and M5 disappears below ~ 150K respectively. To further investigate the temperature dependent behavior of these modes i.e., frequency, FWHM and intensity is shown in Fig. 4 where solid blue line is a guide to the eye. Interestingly most of these emergent modes (M1-M4) show an asymmetry in the line shape i.e., Fano line shape, which suggests presence of interaction of discrete state (phonons) with an underlying continuum possibly electronic in nature [56]. We have qualitatively analyzed and discussed the Fano asymmetry of these phonon modes using a slope method and the temperature evolution of asymmetry is plotted in Figure S7 in section SIII of S.I. [47].

Frequency of M1 and M5 remains nearly constant throughout. M2 mode shows a linear increase with decrease in temperature till ~200K and remains nearly independent of temperature on further decreasing the temperature till 100K. M3 frequency increases marginally from 320K to ~150K and then redshifts continuously till 100K, whereas M4 continuously blueshifts till 150K.

FWHM of M2, M3 and M5 demonstrates nearly temperature independent behavior throughout. M1 shows a continuous decrease in linewidth with decrease in temperature till 100K, whereas for M4 it decreases continuously ~220K and remains constant below till ~150K. As far as temperature dependent variation of intensity is concerned. M2 and M5 mode exhibits nearly constant behavior. M1, M3 and M4 shows expected behavior i.e., decreasing with decrease in temperature.

### 2.4 Electron-phonon coupling (EPC)

As electrons move through the crystal lattice, they can disperse off lattice vibrations, leading to a change in momentum and energy of both the electron and the phonon. In metals for example, low-energy excitations are significantly modified due to interaction with the lattice vibrations which consequently impacts the transport and thermodynamic behavior. In the context of superconductivity, strong electron-phonon coupling leads to the formation of Cooper pairs. In case of electron-phonon coupling the scattering, probability is dictated by Fermi's golden rule, where a phonon excites an electron into higher momentum state.

Temperature dependent Raman scattering is a sensitive technique to detect the presence of electron-phonon scattering. Generally, phonon mode frequency blueshifts with decrease in temperature due to a combined effect of lattice expansion and anharmonic effects [52,57]. The observation of anomalous redshift in frequency for P2 and P3 mode below ~ 100K can be construed as evidence of presence of electron phonon coupling due to intra-band fluctuations in this semimetal [58]. The broadening in the phonon linewidth is related to the phonon life-time.

Normally it arises from anharmonic decay of optical phonons to acoustic phonons and the linewidth is expected to scale with the Bose factor $n_B(\omega,T)$ i.e., decreases with decreasing temperatures. Its value depends on interaction of phonon with various quasi-particles for e.g., phonon-phonon, electron-phonon interaction etc. Non-monotonic behavior of FWHM of P3, P4 and P5 modes also suggests a strong electron-phonon coupling [59]. Such an anomalous nature of FWHM can be captured by a model based on phonon decaying into electron-hole pairs. In this case linewidth behavior is dictated by Fermi-Dirac occupation factor, $n_F(\omega,T)$. Such a phenomenological expression for linewidth is given as follows [38,60]:

$$\Gamma(T) = \Gamma_o + \eta \ [n_F(\hbar\omega_a,T) - n_F\{\hbar(\omega_a+\omega_p),T\}] \qquad -- (7)$$

here $\Gamma_o$ is a temperature independent term which incorporates any of anharmonic or electron-phonon contribution to linewidth at absolute zero. $\eta$ is an indicator of coupling strength and the available density of states for decay [61]. $\omega_p$ is the optical phonon frequency whereas $\omega_a$ is the energy difference between the electron's initial state and the Fermi energy in the phonon-mediated inter-band scattering. We have fitted the FWHM of P3 and P4 modes using eq. 7 as shown in Fig. 3 (Olive solid line). Phenomenologically it can be understood as, when $\omega_a \approx \omega_p$ it corresponds to the case where phonon decays into an electron-hole pair. As temperature increases phonon no longer decompose into an electron-hole pair as the final state (hole) is already occupied hence the linewidth decreases with increase in temperature. If $\omega_a \approx 0$, it indicates that the electronic states are already occupied at $T = 0K$, hence linewidth decreases monotonously with increasing temperature [60]. That is what observed in case of P3, P4 and P5 mode and the obtained value of $\omega_a$ is very close to zero. So, we find that eq. 7 dictates well the behavior of linewidth through-out the temperature range.

## 2.5 Polarization dependent phonons analysis

To unveil symmetry of the phonon modes we performed an angle resolved polarized Raman scattering experiment in a configuration where we controlled the direction of polarization of the incident light ranging from $0^o$ to $360^o$ using a half wave retarder ($\lambda/2$ plate), whereas analyzer principal axis has been kept fixed. The experiment is performed at two different temperatures i.e., room temperature (~300K) and 14K and angular variation of intensities of modes are shown in Fig 5. Within semiclassical approach, the Raman scattering intensity from first-order phonon modes is related to the Raman tensor, polarization configuration of incident and scattered light which can be defined as:

$$I_{Raman} = C\left[\sum_{k,l=x,y,z} e_i^k R_{kl} e_s^l\right]^2 = C\left|\hat{e}_s^\dagger . R . \hat{e}_i\right|^2 \quad\quad\quad \text{--- (8)}$$

where '$\dagger$' symbolizes transpose, $'\hat{e}_i'$ and $'\hat{e}_s'$ are the Jones vectors representing incident and scattered light polarization direction. '$R$' represents the Raman tensor of the respective phonon mode [62-65]. A schematic diagram of polarization vectors of incident and scattered light projected on a plane is shown in Fig. 1(c). In the matrix form, incident and scattered light polarization direction vector can be decomposed as: $\hat{e}_i = [\cos(\alpha+\beta)\ \sin(\alpha+\beta)\ 0]$ ; $\hat{e}_s = [\cos(\alpha)\ \sin(\alpha)\ 0]$, where '$\beta$' is the relative angle between $'\hat{e}_i'$ and $'\hat{e}_s'$ and '$\alpha$' is an angle of scattered light from x-axis, when polarization unit vectors are projected in x (*a*-axis) - y (*b*-axis) plane. The Raman tensors for various modes are summarized in Table-I. The angular dependency of intensities of the Raman active modes using eq. 8 can be written as:

$$I_{E'(x)} = |d\ \cos(2\alpha+\beta)|^2,\ I_{E'(y)} = |d\ \sin(2\alpha+\beta)|^2,\ I_{E''} = 0 \text{ and } I_{A_1'} = |a\ \cos(\beta)|^2 \quad \text{--- (9)}$$

Here $\alpha$ is an arbitrary angle from the *a*-axis and is kept constant. Therefore, without any loss of generality it can be taken as zero, giving rise to the expression for the Raman intensity as $I_{E'(x)} = |\text{d} \cos(\beta)|^2$ and $I_{E'(y)} = |\text{d} \sin(\beta)|^2$.

Angular dependence of the intensity for P2-P5 and M1-M5 is shown in Fig. 5 at 14K and room temperature (~ 300K). Fitted spectra at different angles for room temperature is shown in Fig. S2 (S.I.) [47] . First, we will discuss about the prominent modes i.e., P2-P5. We can see that at room temperature P2, P3, P4 shows $E'(x)$ or $A'_1$ kind of behavior whereas P5 shows elliptical type behavior. While at 14K, P2 mode exhibits nearly similar behavior as compared to that at room temperature. But one can notice a rotation in major axis by ~10º, area in the 3$^{rd}$ and 4$^{th}$ quadrant has decreased. At room temperature maxima of P3 is observed at ~10º but at 14K it has shifted to 30º/160º. Whereas P4 and P5 shows very distinct variation of intensity as compared to the room temperature. Change in the symmetry of the modes at different temperatures may be owing to a possible structural phase transition. Further, interestingly at room temperature mode M1- M5 appears only at certain angles, discussed in detail in the next section.

## 3. Discussion

We will be focusing here only the possible origin of weak modes M1-M5, which appears in the vicinity of 100K- 150K. Kung et al., in their Polarization Resolved Raman Scattering and DFT based study on a topological insulator $Bi_2Se_3$ has reported presence of four additional modes with lower intensity and energy than the bulk counterparts [66]. These modes appear due to c-axis lattice distortion and surface van der Waals gap expansion where these modes become Raman active due to reduction of crystalline symmetry from $D_{3d}$ in bulk to $C_{3v}$ on the surface. Additionally, these modes were observed to have an asymmetry in the line shape i.e., Fano asymmetry , suggesting interference of electron-phonon coupling with the surface excitations [66]. Here a finite phonon density of states exists across the entire BZ and because of that surface modes decay into bulk phonons which reflects that surface is not entirely separated from the bulk. A surface resonance can be expected with lower energy profile than the bulk [66,67]. Concept of surface phonons was initially developed by Lifshitz and Rosenzweig [68-70] where the symmetry of the crystal breaks when we move from bulk to the surface. These modes are localized at the surface where dispersion can be very different than the bulk and phonon frequencies at the surface gets consequently modified for $\Gamma$-point [70,71]. It can be considered as a perturbation of an infinite lattice to derive the surface modes from the bulk mode spectrum. Generally, these modes are only modified to a minimal value as compared to the bulk. Though presence of gap in phonon density of states with large distortion can separate these modes from the bulk [72]. In case of topological insulator $Bi_2Se_3$ surface effects are also reflected in both bulk and surface electronic band structure [67,73,74].

Recently, Murtaza et al., reported a series of pressure-driven phase transitions in $PbTaSe_2$ upto 56 GPa, with space group $P\bar{6}m2$ (α-phase) at ~ 0 GPa, $P6_3mc$ (β-phase) at ~0.5 GPa [21]. The $P6_3mc$ structure can be obtained by doubling the unit cell of the $P\bar{6}m2$ structure along the c-axis and relocating the upper half of the unit cell by 1/3 along the diagonal of the basal plane [1/3(b-a)] [19]. Corresponding group theoretical prediction of Raman active modes for different phases are mentioned in table SIII (S.I.). In the α-phase there are two bands crossing each near the Fermi level where it forms two nodal lines on the plane of $k_z = \pi/c_\alpha$ (where $c_\alpha$ is the c lattice constant for α-phase, etc.). Whereas for β-phase multiple types of band crossing has been observed, including type-II Weyl nodal lines, type-II Dirac points, and twofold nodal surface, near the Fermi level [21]. There is a twofold nodal surface located on the plane of $k_z = \pi/c_\beta$ as a protection of $S_{2z}$ and T [75,76].

Generally, on decreasing the temperature a CDW phase is accompanied by the appearance of new Raman active modes which are denoted as amplitude or phase mode and are distinguished based on their dispersion characteristics [28,40,77,78]. Study of such modes in Raman spectra helps to comprehend the stability of CDW phases [79-81]. The disappearance of modes M1-M5 on decreasing the temperature is unlike a CDW-like transition. In addition to that the presence of Fano asymmetry in modes M1-M4 is a signature of electron-phonon coupling with underlying electronic continuum of the same symmetry which is crucial to understand the relaxation and scattering of surface state excitations. We observed an increase in asymmetry of modes M1-M4 with increasing temperature as shown in Figure S7 of S.I.. Such a behavior is expected as the topological nature is active in this phase. Increasing asymmetry corresponds to the increasing electron-phonon coupling with the topological surface excitations with increasing temperatures and these mode gains more intensity consequently. More details are discussed in S.I. section SIII [47]. The fact that caught our attention is the appearance modes i.e., M1-M5 and the discontinuities present in P2-P5 modes i.e., in frequencies, FWHM and Intensities, occurs around the similar temperature range.

Based on the above different possible scenario, discussed above and our Raman spectroscopic analysis, we assign these weak modes as surface phonon modes. Generally increasing pressure [82,83] or decreasing the temperature [52,84] causes a similar effect as in both case a decrease in the bond length is observed which causes a blueshift in the phonon modes frequencies [85,86]. So, there might be a possibility that as we decrease the temperature it undergoes from α to β phase at $T_{\alpha \to \beta}$ ~ 150 K. Due to this transition the surface phonon spectrum is being modified because in the β-phase, the surface states are affected considerably as discussed above and consequently, we observe disappearance of these weak modes M4-M5. Further we observe another transition $T_{CDW+\beta}$ ~ 100K, below this temperature there is CCDW phase reported for 1H-TaSe$_2$. Such a transition in PbTaSe$_2$ could possibly be due to the interplay of underlying CCDW phase and surface topology of the material because of which M1-M3 also disappear. The possibility of structural transition is also supported by our SC-XRD measurements, where we find appearance of new XRD peaks on decreasing temperatures. Interestingly mode M1-M5 appeared only at certain angles in our polarization dependent analysis at room temperature. There might be a possibility that these modes are getting activated only at certain angles depending on the crystal symmetry at surface and the topological nature of the states in a particular phase which couples with the surface lattice dynamics.

## 4. Conclusion

In conclusion, we conducted a temperature dependent SC-XRD to unveil the structural transition and an inelastic light scattering Raman measurements (temperature and polarization dependent) on single crystals of PbTaSe$_2$. We observed discontinuities in temperature evolution of frequency, FWHM and Intensity for prominent phonon modes which are consistent throughout the temperature range i.e., 6K – 320K (P2-P5). The polarization dependent results for these modes shows a change in symmetry at different temperature (14K and 300K), which suggests a structural change. We observed disappearance of some weak modes on decreasing temperature i.e., M1-M5. These modes vanish at around the temperatures where the discontinuities in prominent phonon modes arises i.e., ~100K (M1-M3) and ~150K (M4-M5). Most of these modes (M1-M4) exhibits Fano asymmetry in the line shape which is indication of presence of electron-phonon coupling. Further at room temperature the polarization dependence analysis of these modes reveals that M1-M5 modes appear at certain angles only. Hence based on our SC-XRD and Raman analysis we suggest that there is a thermally induced structural phase transition happening which is possibly causing it to undergo from α to β phase ~ 150K. The transition ~100K could be owing to the interplay of underlying CCDW phase of 1H-TaSe$_2$ and β phase. We propose the modes (M1-M5) which are disappearing on decreasing temperature are surface phonon modes. Instead of speculations, however the precise understanding of the proposed surface phonon modes still needs to be understood. So, we would like to keep this an open problem for condense matter community for now and hope that our result will motivate more research groups for further theoretical and experimental investigation.

## Acknowledgments

P.K. thanks SERB (Project no. CRG/2023/002069) for the financial support and IIT Mandi for the experimental facilities.

**Figures**

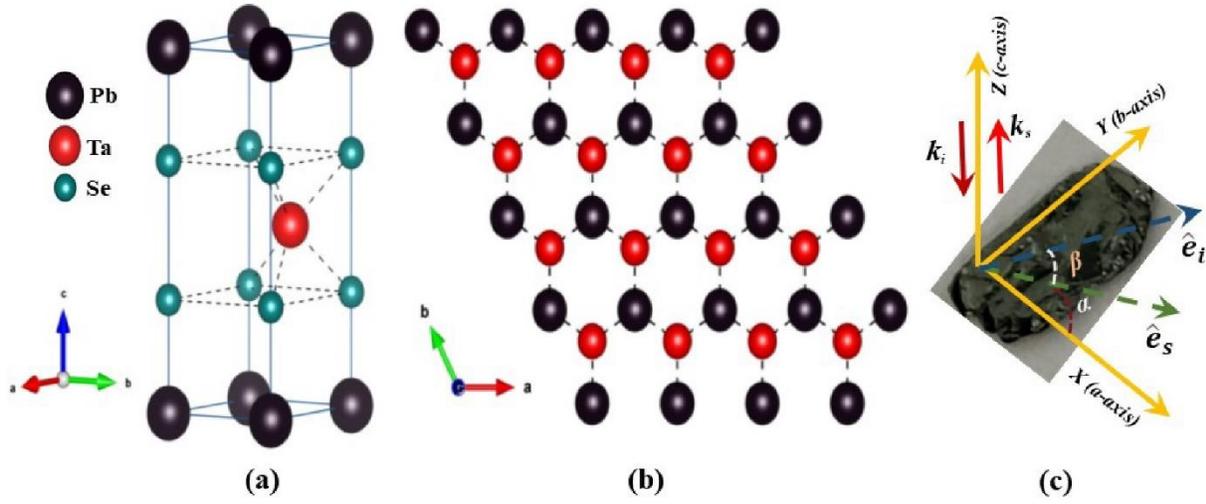

**Figure 1:** Shows structure of the PbTaSe$_2$ in **(a)** standard orientation **(b)** along c-axis and **(c)** Plane projection of polarization direction of incident and scattered light.

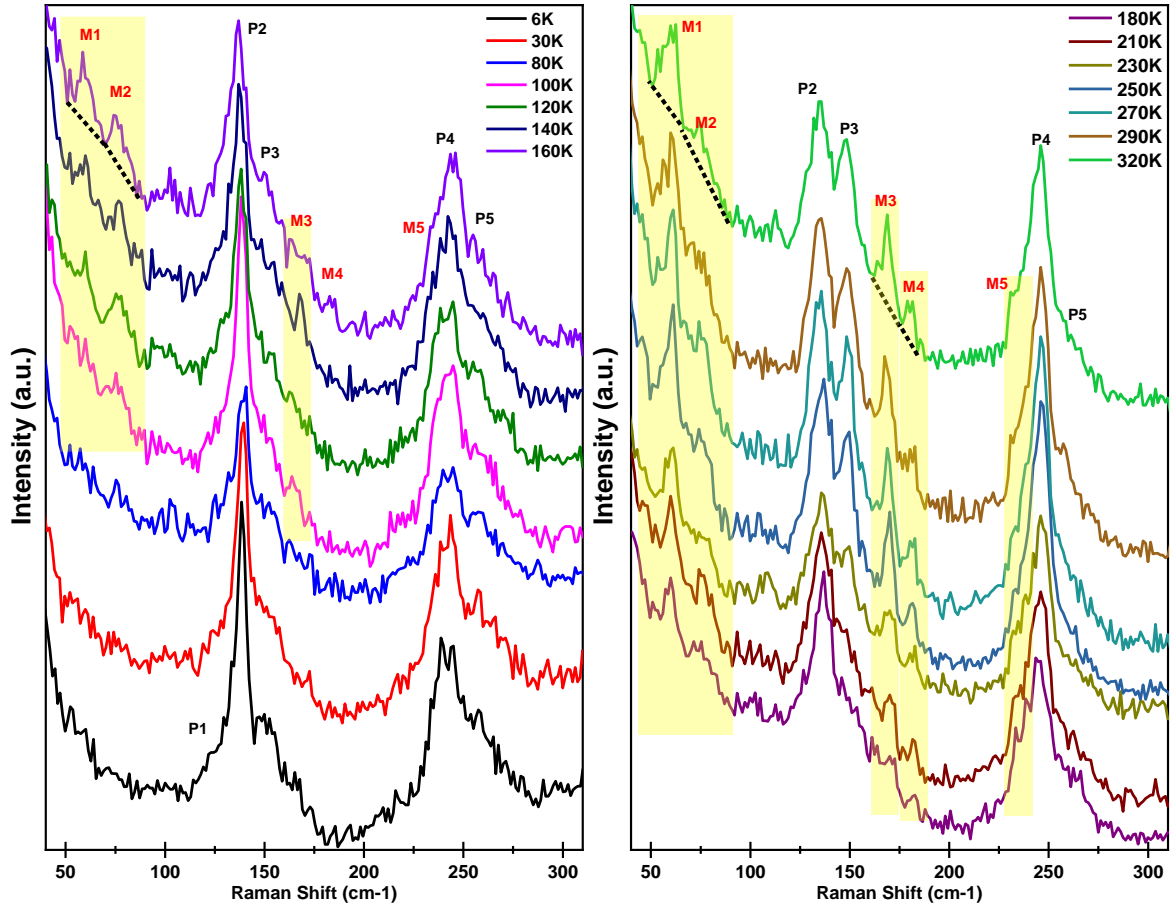

**Figure 2:** Temperature evolution of the Raman Spectra. Prominent modes labeled as P1-P5. Shaded yellow region shows emergence of new modes on increasing temperatures i.e., M1-M5 and dotted black line shows the asymmetry in the line shape of M1-M4 modes.

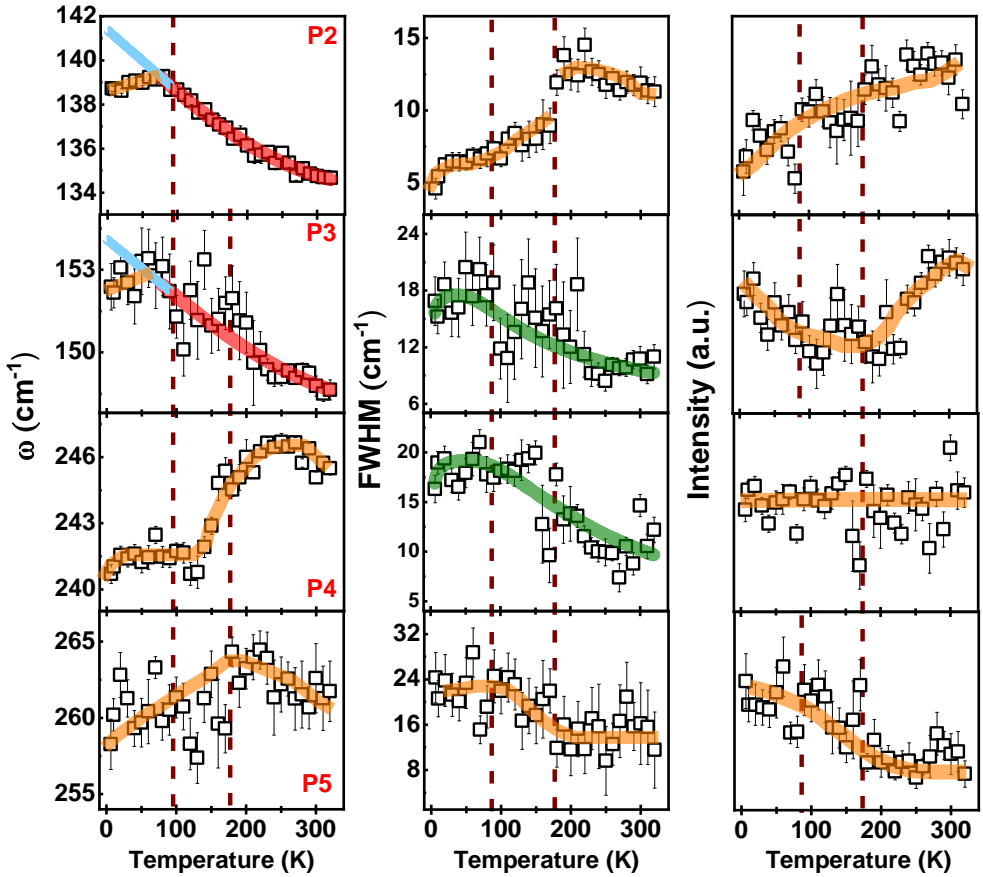

**Figure 3:** Temperature dependent evolution (frequency, FWHM and intensity) of the P2-P5 phonon modes. Red, light blue solid line shows the anharmonic fitting, linear extrapolation, respectively for frequency. Orange solid line is a guide to the eye. Solid Olive line is a fit for FWHM of P3 and P4, with a model mentioned in the text. Dashed line shows the position of discontinuities.

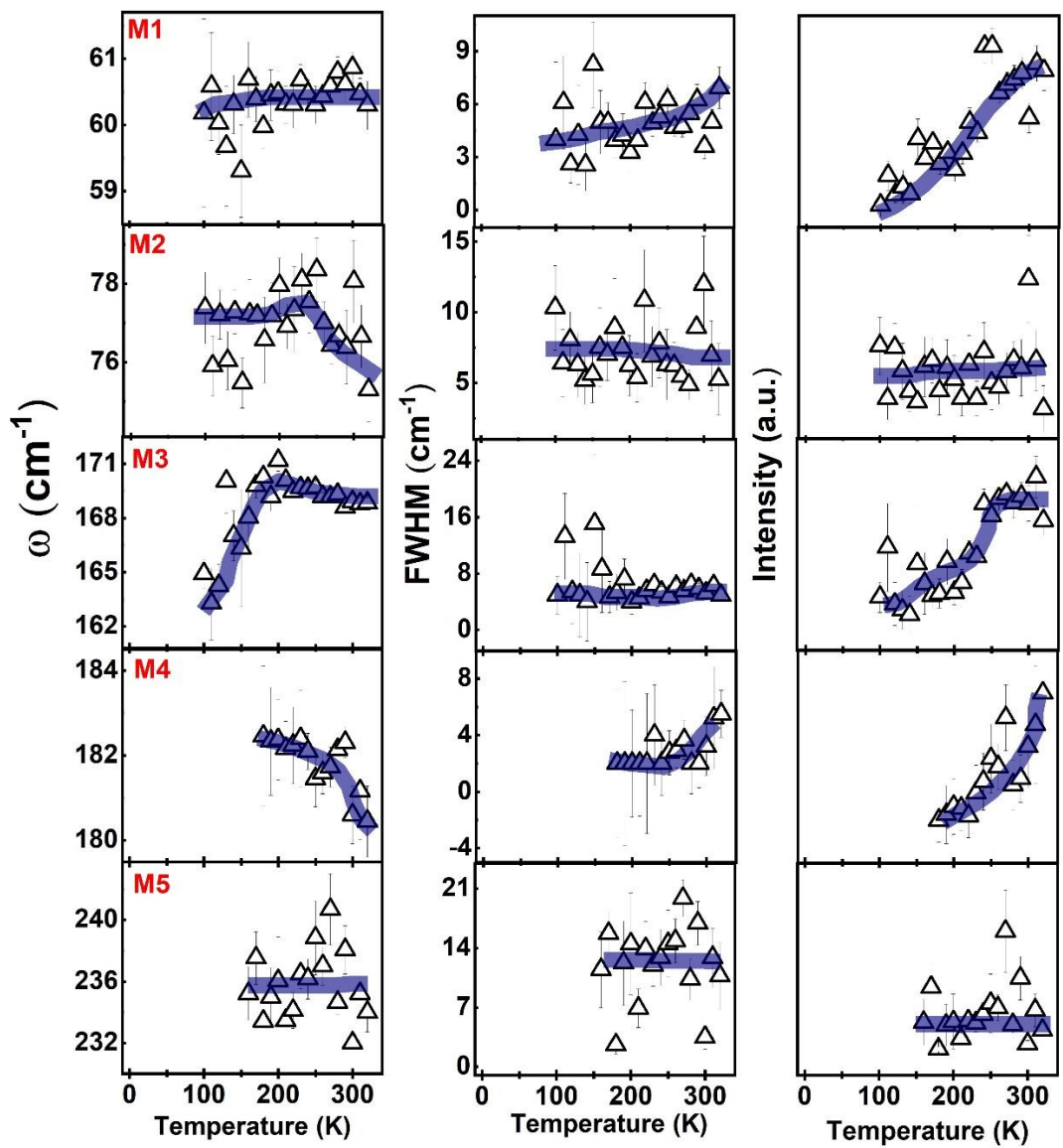

**Figure 4:** Temperature dependent evolution (frequency, FWHM and intensity) of the M1-M5 phonon modes. Blue solid line is a guide to eye.

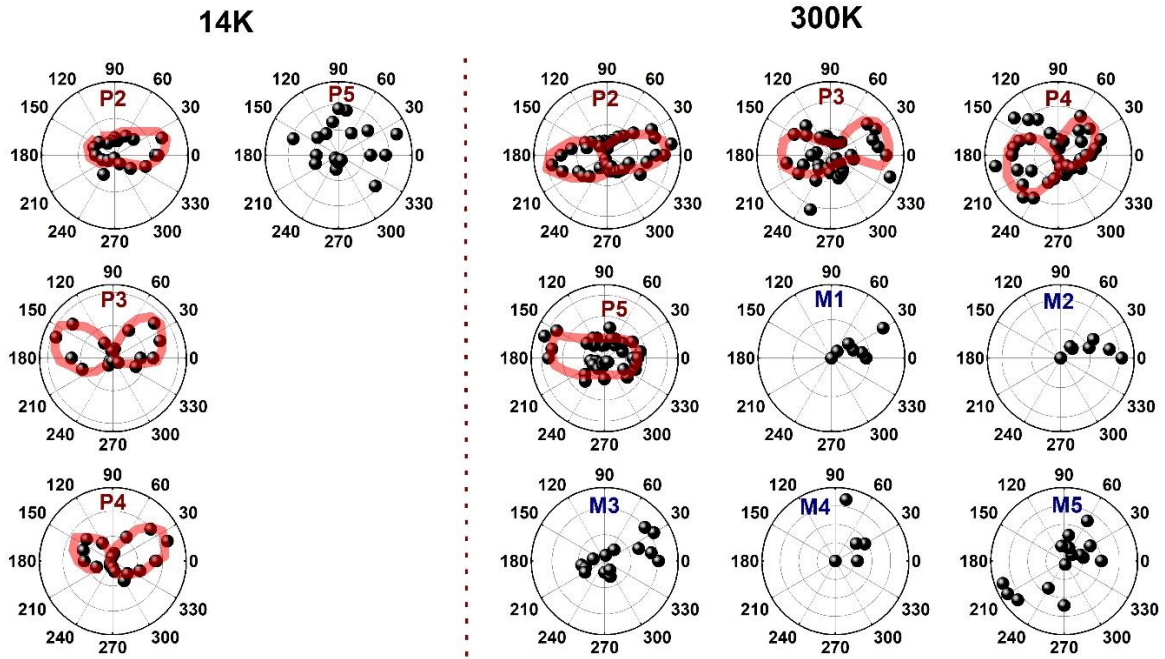

**Figure 5:** Polarization dependent intensity variation of the observed phonon modes at 14K and 300K. Red solid line is guide to the eye.

**Table I:** Wyckoff positions of different atoms in conventional unit cell and irreducible representations of the phonon modes of hexagonal (#187; $P\bar{6}m2$ [$D_{3h}$] PbTaSe$_2$) at the gamma point, along with Raman Tensors of the Raman active phonon modes.

| Atoms | Wyckoff site | Γ-point mode decomposition | Raman Tensors |
|---|---|---|---|
| Pb | 1a | $A_2''$ (I.R) + $E'$ (I.R.+R) | $E'(x) = \begin{pmatrix} d & 0 & 0 \\ 0 & -d & 0 \\ 0 & 0 & 0 \end{pmatrix}$; $E'(y) = \begin{pmatrix} 0 & -d & 0 \\ -d & 0 & 0 \\ 0 & 0 & 0 \end{pmatrix}$ |
| Ta | 1d | $A_2''$ (I.R) + $E'$ (I.R.+R) | $E'' = \begin{pmatrix} 0 & 0 & 0 \\ 0 & 0 & c \\ 0 & c & 0 \end{pmatrix}$; $E'' = \begin{pmatrix} 0 & 0 & -c \\ 0 & 0 & 0 \\ -c & 0 & 0 \end{pmatrix}$ |
| Se | 2g | $A_1'(R)$ + $A_2''$ (I.R.) + $E'$(I.R.+R) + $E''(R)$ | $A_1' = \begin{pmatrix} a & 0 & 0 \\ 0 & a & 0 \\ 0 & 0 & b \end{pmatrix}$ |

$\Gamma_{\text{Raman}} = A_1' + 3E' + E''$    $\Gamma_{\text{Infrared}} = 3A_2'' + 3E'$

**Table II:** Experimentally observed modes and anharmonic fit parameter. Units are in cm$^{-1}$.

| Modes | $\omega_{EXP.}$ | $\omega_o$ | A/B |
|---|---|---|---|
| M1 (100K) | 60.2 ± 1.4 | | |
| M2 (100K) | 77.4 ± 0.9 | | |
| P1 (6K) | 133.4 ± 2.5 | | |
| P2 (6K) | 138.7 ± 0.1 | 142.5 ± 0.3 | -26.4 |
| P3 (6K) | 152.4 ± 0.8 | 155.0 ± 1.3 | -28.3 |
| M3 (100K) | 164.9 ± 0.8 | | |
| M4 (150K) | 182.5 ± 1.6 | | |
| M5 (150K) | 235.2 ± 1.7 | | |
| P4 (6K) | 240.7 ± 0.4 | | |
| P5 (6K) | 258.3 ± 1.6 | | |

**Surface phonons and possible structural phase transition in a topological semimetal PbTaSe$_2$**

Vivek Kumar[1,*] and Pradeep Kumar[1,†]

[1] School of Physical Sciences, Indian Institute of Technology Mandi, Mandi-175005, India

*vivekvke@gmail.com
†pkumar@iitmandi.ac.in

## Supplementary Information

**S1: Effect of thermal expansion**

The coefficient of thermal expansion, denoted by $\alpha(T) = 1/V(\partial V/\partial T)$, is used to express volume change due to variation in temperatures and related to the average vibrational amplitude of the phonons [1]. Here, we will discuss the thermal expansion contribution which is to comprehend the role of third term in Eq. 1 in the main text. For that we refer to the Gruneisen constant model which describes the volume expansion effect on modification of the phonon mode frequency and is given by [2]:

$$\delta\omega_E(T) = \omega_E(T) - \omega_o = \omega_o \exp[-3\gamma \int_{T_o}^{T} \alpha(T)dT] - \omega_o \qquad ---(1)$$

Here $\gamma$ is the Gruneisen parameter and $\alpha(T)$ is the temperature dependence of thermal expansion coefficient (TEC). The integral depicts redshift of phonon because of thermal expansion. The product of the Gruneisen parameter and the TEC can be expressed in the form of a polynomial as:

$$\gamma * \alpha(T) = a + bT + cT^2 \qquad -- (2)$$

where a, b and c are the parameters which are obtained by fitting. We have used value of Gruneisen parameter $\gamma = 2.24$ as reported in a DFT based research by Reza et al., which suggests a moderate level of anharmonic effect in this material [3].

We have fitted the frequency of P2 and P3 modes in temperature range of 100K-320K taking both anharmonic and thermal expansion terms and is shown in Fig. S1 where thick red line shows the fit. We estimated the temperature dependence of TEC for these modes as shown in Fig. S1 (b). Interestingly we note that $\alpha(T)$ for P2 and P3 display a negative value below 220K and 120K respectively. The value of $\alpha(T=300K)$ for P3, P4 is 20.2 ($10^{-5}$ $K^{-1}$) and 07.1 ($10^{-5}$ $K^{-1}$) respectively.

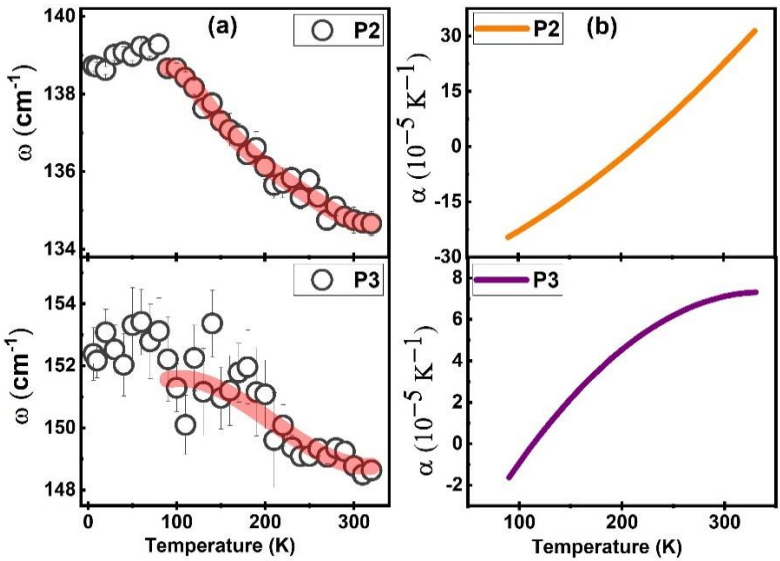

**Figure S1: (a)** Temperature dependent frequency variation of P2 and P3 mode fitted (solid red line) with model as mentioned in the text. **(b)** $\alpha(T)$ for P2 and P3 phonon modes.

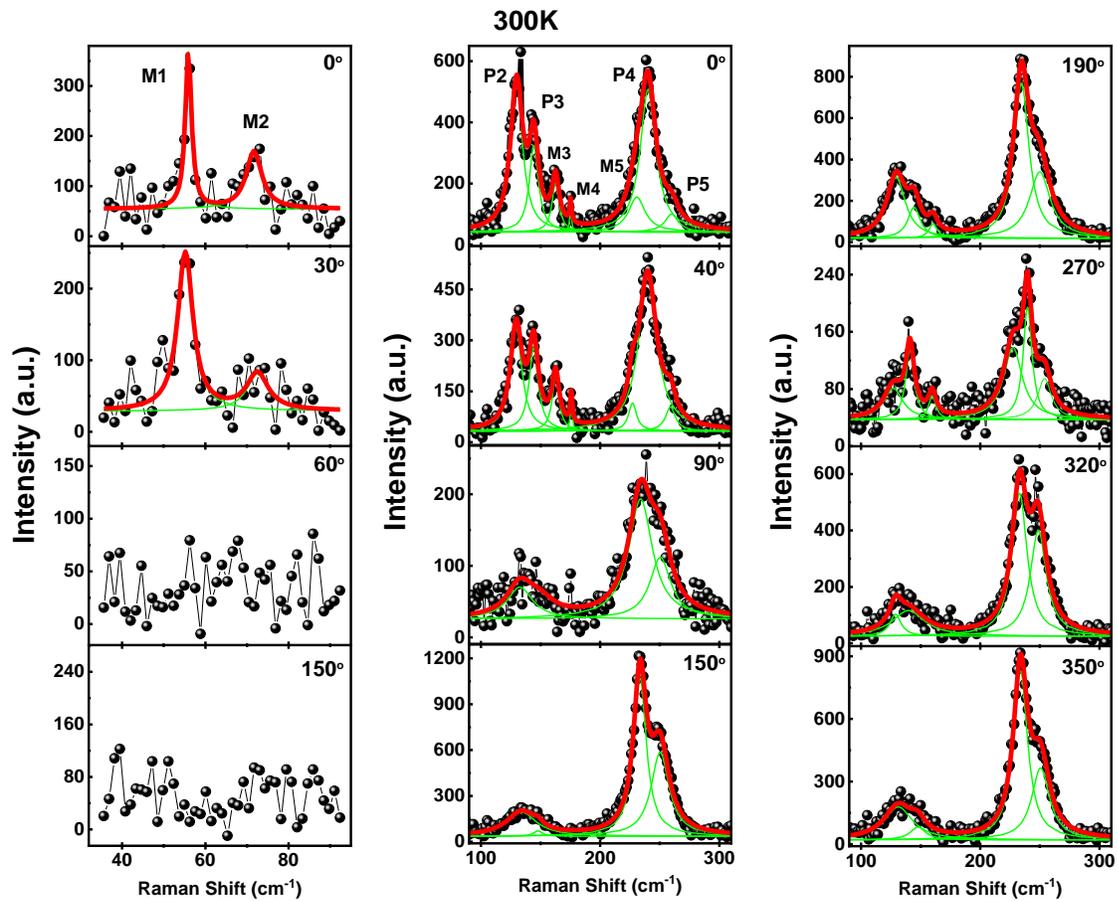

**Figure S2:** Fitted spectra at different polarization angle at 300K.

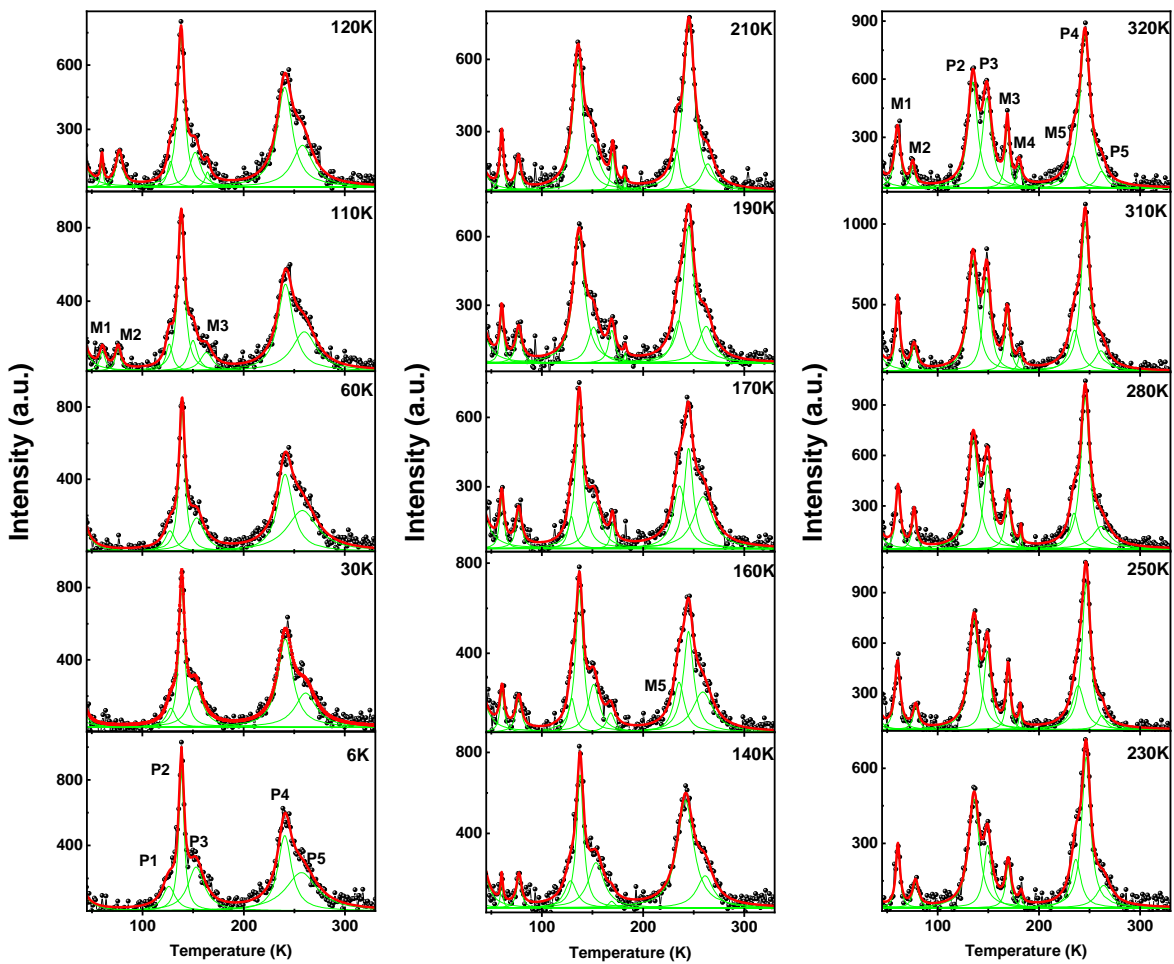

**Figure S3:** Fitted spectra at different temperatures.

**SII: Single crystal X-ray diffraction (SC-XRD) measurements**

To reveal the possibility of temperature dependent (TD) structural transition, we carried out a TD SC-XRD measurement using SuperNova X-ray diffractometer system with a copper (Cu, $\lambda \sim 1.54$ Å) source to radiate the sample at ambient pressure. Liquid nitrogen is utilized to vary temperature from 87K till room temperature. A clear signature of structural transitions is observed as appearance of new peaks on decreasing temperatures which are indicated with '*' symbol in the raw SC-XRD spectra as shown in figure S4 and figure S5 (background is removed). Appearance of new peaks $\sim 27^o$, $39^o$, $45^o$ and $72^o$ occurs below 120K whereas peak $\sim 22^o$, show broadening above $\sim 200$K. The XRD spectra obtained at room temperature is consistent with previous report [4].

We also generated XRD pattern from VESTA using structural parameters reported by Udhara S. Kaluarachchi et al. [5] and have plotted the generated Braggs positions for α and β-phase. For comparison purpose we have plotted it for 87K and room temperature XRD spectra, as shown in figure S6. Our observed data for room temperatures do matches with the α-phase. In case of 87K XRD spectra, we found a deviation from the generated Braggs position for β-phase. Still, we observe that some new peaks $\sim 45^o$ and $\sim 72^o$ which were not present in the α-phase do matches with the generated Braggs positions of β-phase. So, we suspect that it is in proximity of the β-phase if not exactly.

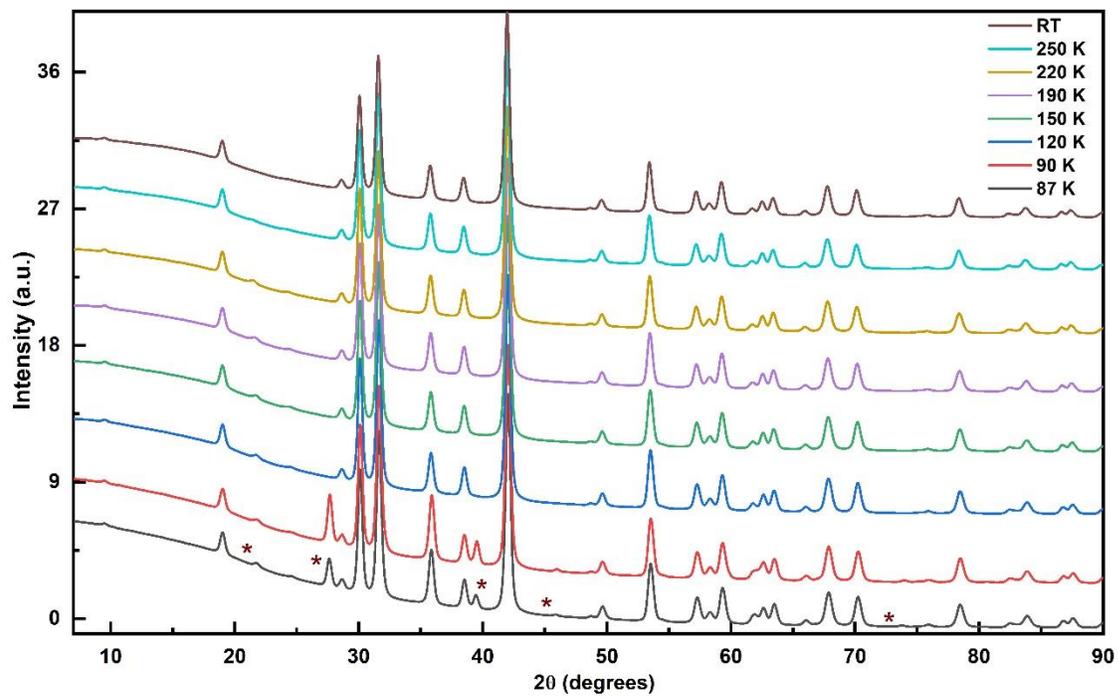

**Figure S4:** Raw SC-XRD spectra at different temperatures. '*' indicates peaks which changes with varying temperatures.

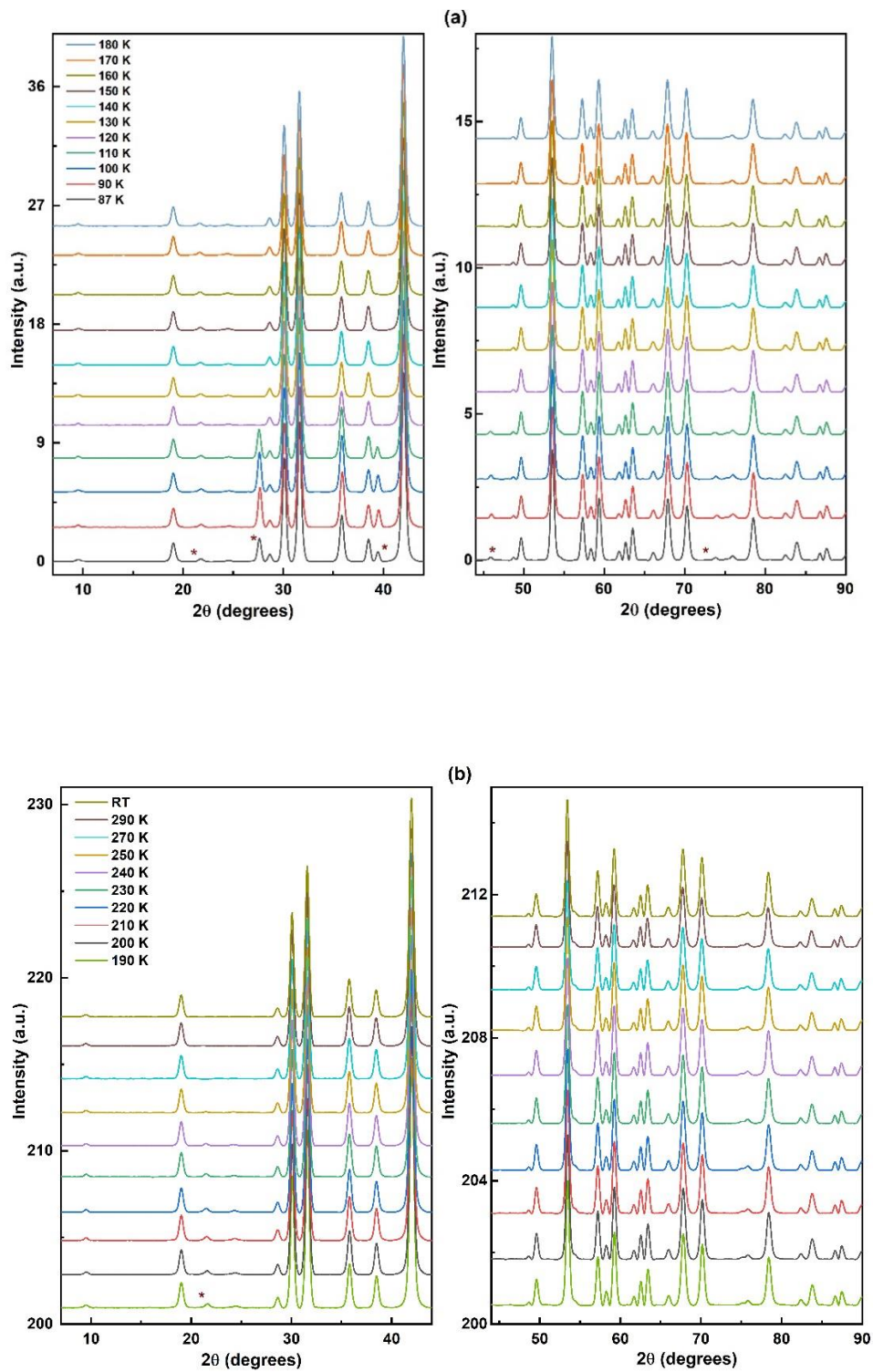

**Figure S5:** (a), (b) shows background subtracted SC-XRD spectra at different temperatures. '*' indicates peaks which changes with varying temperatures.

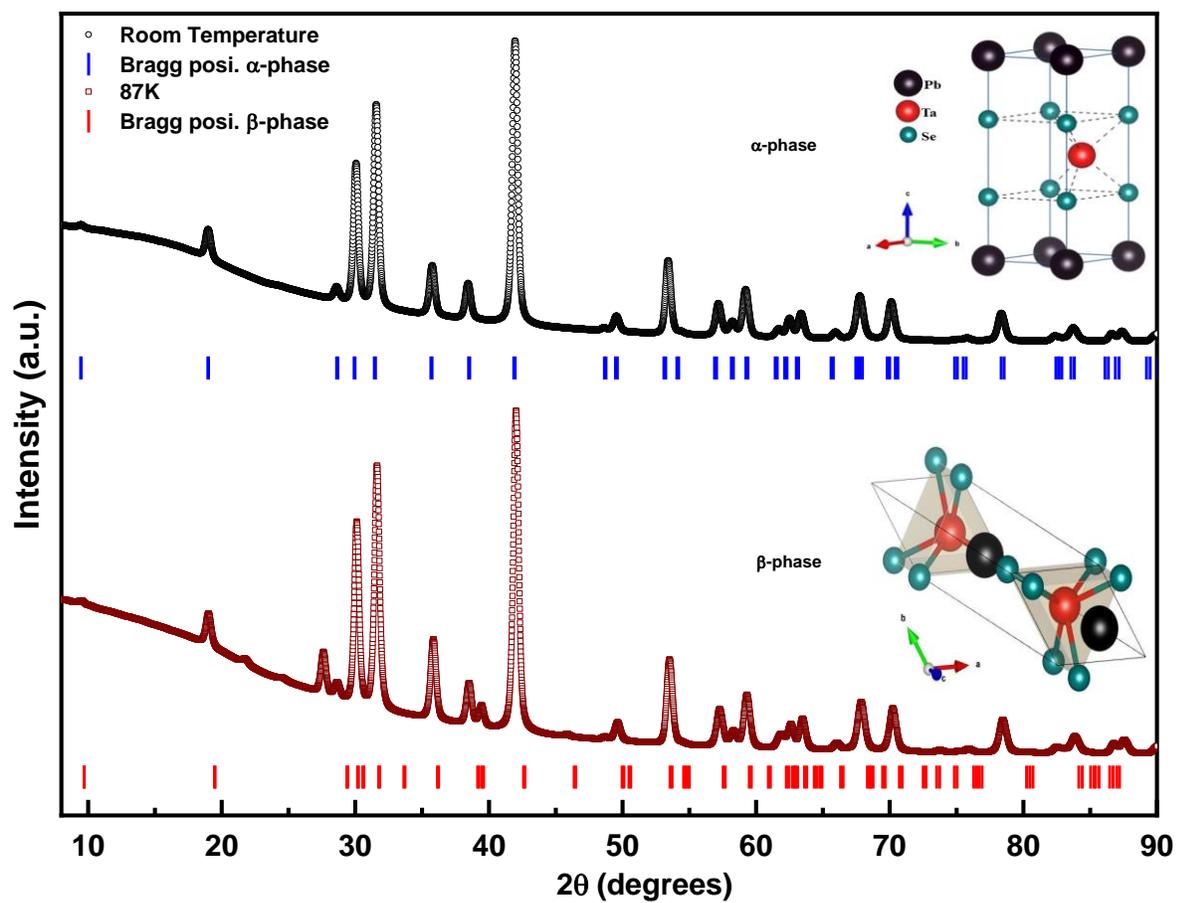

**Figure S6:** Raw SC-XRD spectra at 87K and room temperatures along with Bragg positions generated using VESTA for α and β-phase.

**SIII: Fano asymmetry of M1-M4 phonon modes.**

An equally effective 'Slope method' can be utilized to calculate the qualitative Fano-asymmetric nature of the phonon modes [6]. The slope ($I_2 - I_1 / \omega_2 - \omega_1 = \Delta Inten./\Delta\omega$) for a mode is evaluated by keeping the range of Raman shift same and the corresponding change in the intensity can be plotted as a function of temperature. Fano function can be written as $F(\omega) = I_0(q+\varepsilon)^2/(1+\varepsilon^2)$; where $\varepsilon = (\omega-\omega_0)/\Gamma$ and $1/q$ defines as the asymmetry [7]. The asymmetry parameter ($1/q$) characterizes the coupling strength of a discrete state (phonons) to the underlying continuum. Unlike a symmetric Lorentzian profile, the asymmetry in line shape is determine by $q$, a stronger coupling ($1/q \rightarrow \infty$) causes the peak to be more asymmetric and in the weak-coupling limit ($1/q \rightarrow 0$) where the Fano line shape is reduced to a Lorentzian line shape. there is a one-to-one correspondence between slope and asymmetry parameter $1/q$, higher the slope more is the asymmetry and lesser the slope lesser the asymmetry. So, phenomenologically both depict the same feature. As observed M1-M3 modes appears at ~100K and M4 appears at ~150K. In Figure S7 we have plotted the mode of the slope (solid orange line is guide to eye) in temperature range of 100K-320K for modes M1-M4 for which we observed Fano asymmetry as shown in Figure 1 in the manuscript. On decreasing temperatures slope of mode M1 show linear decrease till ~ 200K before increasing slightly with further decrease in the temperature till 100K. For M2 it decreases monotonously till ~200K and remains nearly constant till 100K. Slope of M3 decreases till ~200K and then remains nearly constant on further decreasing temperatures. Mode M4 slope decreases monotonously till ~150K. We observed an overall increase in asymmetry with increasing temperatures for M1-M4 modes. Such a behavior is expected as the topological nature is active in this phase. Increasing asymmetry corresponds to the increasing electron-phonon coupling with the topological surface excitations with increasing temperatures and these modes gains more intensity consequently.

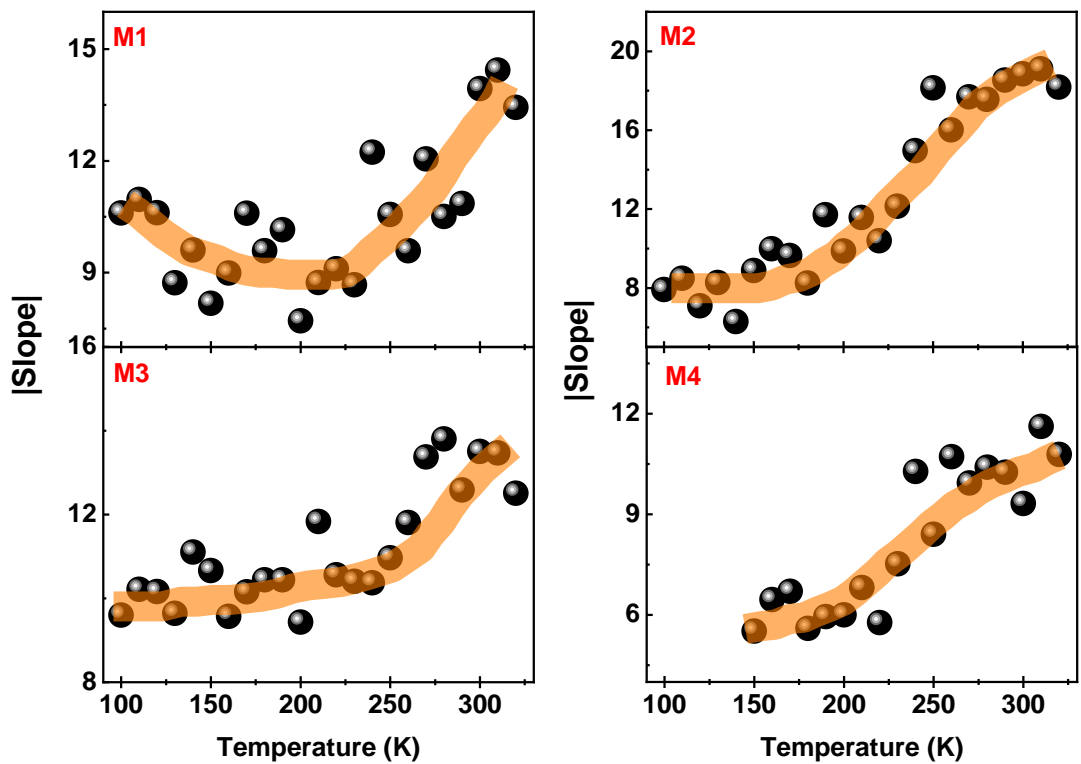

**Figure S7:** Temperature dependence of mode of slope for M1-M4 phonon modes, indicating the Fano -asymmetry. Solid orange line is guide to the eye.

**Character Table**

**Table SI:** Character Table for the space group P$\bar{6}$m2.

| D$_{3h}$(P$\bar{6}$m2) | # | 1 | m | 3 | -6 | 2$_{120}$ | m$_{100}$ | function |
|---|---|---|---|---|---|---|---|---|
| Mult. | - | 1 | 1 | 2 | 2 | 3 | 3 | . |
| $A'_1$ | $\Gamma_1$ | 1 | 1 | 1 | 1 | 1 | 1 | $x^2+y^2, z^2$ |
| $A'_2$ | $\Gamma_2$ | 1 | 1 | 1 | 1 | -1 | -1 | J$_z$ |
| $A''_1$ | $\Gamma_3$ | 1 | -1 | 1 | -1 | 1 | -1 | . |
| $A''_2$ | $\Gamma_4$ | 1 | -1 | 1 | -1 | -1 | 1 | Z |
| $E'$ | $\Gamma_5$ | 2 | 2 | -1 | -1 | 0 | 0 | (x,y),(x$^2$-y$^2$,xy) |
| $E''$ | $\Gamma_6$ | 2 | -2 | -1 | 1 | 0 | 0 | (xz,yz),(J$_x$,J$_y$) |

**Table SII:** DFT Calculation for α-phase (P6̄m2).

| Modes # (Symmetry) | $\omega_{DFT}$ (cm$^{-1}$) | Optical Activity | e-ph (GHz) |
|---|---|---|---|
| 1 ($E'$) | -6.9 | I+R | 0.05 |
| 2 ($E'$) | -6.9 | I+R | 0.05 |
| 3 ($A_2''$) | 4.9 | I | 0.05 |
| 4 ($E'$) | 23.2 | I+R | 0.17 |
| 5 ($E'$) | 23.2 | I+R | 0.17 |
| 6 ($A_2''$) | 63.9 | I | 0.51 |
| 7 ($E''$) | 150.3 | R | 1.40 |
| 8 ($E''$) | 150.3 | R | 1.40 |
| 9 ($E'$) | 221.1 | I+R | 4.84 |
| 10 ($E'$) | 221.1 | I+R | 4.84 |
| 11 ($A_1'$) | 228.3 | R | 4.80 |
| 12 ($A_2''$) | 272.9 | I | 2.66 |

**Table SIII:** Space group, point group and Raman active modes in different phases.

| Phase | Space group | Point group | Raman active modes |
|-------|-------------|-------------|--------------------|
| α | P$\bar{6}$m2 | D$_{3h}$ | $A_1' + 3E' + E''$ [8,9] |
| β | P6$_3$mc | C$_{6v}$ | $3A_1 + 4E_2 + 3E_1$ [10] |